\documentclass[aps,prl,twocolumn,superscriptaddress]{revtex4-1}
\usepackage{graphicx}
\usepackage{latexsym}
\usepackage{amssymb}
\usepackage{amsmath}
\usepackage{amsfonts}
\usepackage{upgreek}
\usepackage{bm}
\usepackage{multirow}
\usepackage{color}
\usepackage{hyperref}
\hypersetup{
colorlinks = true,
linkcolor = [rgb]{0.70,0.13,0.13},
citecolor = [rgb]{0.13,0.55,0.13},
urlcolor = [rgb]{0.25, 0.41, 0.88}}

\newcommand{\ket}[1]{\left|#1 \rangle \right.}

\newcommand{\dsZ}{\mathbb{Z}}

\newcommand{\eqnref}[1]{Eq.\,\eqref{#1}}
\newcommand{\figref}[1]{Fig.\,\ref{#1}}

\usepackage{pdfpages} 
\makeatletter
\AtBeginDocument{\let\LS@rot\@undefined}
\makeatother

\begin{document}

\title{Topological and symmetry-enriched random quantum critical points}

\author{Carlos M. Duque}
\thanks{CD and HH contributed equally to this work.}
\affiliation{Department of Physics, University of Massachusetts, Amherst, Massachusetts 01003, USA}

\author{Hong-Ye Hu}
\thanks{CD and HH contributed equally to this work.}
\affiliation{Department of Physics, University of California, San Diego, CA 92093, USA}

\author{Yi-Zhuang You}
\affiliation{Department of Physics, University of California, San Diego, CA 92093, USA}

\author{Vedika Khemani}
\affiliation{Department of Physics, Stanford University, Stanford, CA 94305, USA}

\author{Ruben Verresen}
\affiliation{Department of Physics, Harvard University, Cambridge, MA 02138, USA}

\author{Romain Vasseur}
\affiliation{Department of Physics, University of Massachusetts, Amherst, Massachusetts 01003, USA}

 \begin{abstract}

We study how symmetry can enrich strong-randomness quantum critical points and phases, and lead to robust topological edge modes coexisting with critical bulk fluctuations. These are the disordered analogues of gapless topological phases. Using real-space and density matrix renormalization group approaches, we analyze the boundary and bulk critical behavior of such symmetry-enriched random quantum spin chains. We uncover a new class of symmetry-enriched infinite randomness fixed points: while local bulk properties are indistinguishable from conventional random singlet phases, nonlocal observables and boundary critical behavior are controlled by a different renormalization group fixed point. We also illustrate how such new quantum critical points emerge naturally in Floquet systems. 

\end{abstract}

\maketitle

Topological phases form a cornerstone of modern condensed matter physics, extending beyond the Landau-Ginzburg paradigm of symmetry-breaking order. An especially important class of topological states are Symmetry-Protected Topological (SPT) phases~\cite{PhysRevB.80.155131,PhysRevB.84.235128,PhysRevB.83.075102,PhysRevB.83.075103,Chen2011b,Pollmann2012,YuanMing2012,Levin2012,Chen1604,Chen2011,doi:10.1146/annurev-conmatphys-031214-014740}, which are gapped systems characterized by non-local order parameters and symmetry-protected topological edge modes. Prominent examples of SPT phases include fermionic topological insulators~\cite{Volkov85,Salomaa88,Murakami04,Kane05,Fu07,Kitaev09,Schnyder09,Hasan10}---protected by time-reversal and charge conservation symmetry---or the Haldane phase in quantum spin chains~\cite{Haldane83,Affleck88,Darriet93,Pollmann10}---protected by spin-rotation symmetry. 

Recently, the concept of SPT order was extended to gapless systems~\cite{Kestner11,Fidkowski11,Sau11,Tsvelik11,Cheng11,Ruhman12,Grover12,Kraus13,Ortiz14,Ruhman15,Iemini15,Lang15,Keselman15,Kainaris15,Zhang16,Ortiz16,Montorsi17,Wang17,Kane17,Ruhman17,Kainaris17,Scaffidi2017Gapless,Guther17,Chen18,Verresen2018Topology,Zhang18,Jiang18,Parker2018Topological,Keselman18,Jones19,Verresen2019Gapless,2019arXiv190901425J,Verresen2020Topology}: surprisingly, many of the key features of SPT physics carry over to the gapless case, despite the non-trivial coupling between topological edge modes and bulk critical fluctuations. It is also helpful to think of gapless SPT (gSPT) states~\cite{Scaffidi2017Gapless} as symmetry-enriched quantum critical points (SEQCP)~\cite{Verresen2019Gapless}, where global symmetries can enrich the critical behavior of critical systems. This led to the discovery of new critical points and phases with unusual nonlocal scaling operators which imply an anomalous surface critical behavior, and symmetry-protected topological edge modes. In certain cases, such SEQCPs are naturally realized as phase transitions separating SPT and symmetry-broken phases: while the bulk universality class is locally dictated by the Landau-Ginzburg theory of spontaneous symmetry-breaking, the nonlocal operators and the surface critical behavior are affected by the neighboring SPT phase.

In this work, we show that the mechanism protecting gapless SPT phases persists upon adding disorder. We focus on one-dimensional systems, where the bulk criticality flows to infinite-randomness fixed points~\cite{Ma1979Random, Dasgupta1980Low-temperature,Fisher1992Random, Fisher1994Random, Fisher1995Critical, Motrunich2000Infinite-randomness}. We first discuss the paradigmatic infinite-randomness Ising criticality, where we find that---similar to the clean case \cite{Verresen2019Gapless}---there are topologically distinct versions in the presence of an additional $\mathbb Z_2^T$ symmetry. We find that one of these classes has topologically-protected edge states. Whilst this is a finite-tuned critical point, our second example is a stable  ``random singlet'' phase of matter. Moreover, in the latter case, there are additional gapped degrees of freedom which are able to make the edge mode exponentially-localized. We also illustrate how this topological random quantum criticality can emerge naturally in periodically driven (Floquet) systems. 

\begin{figure}[t!]
\begin{center}
\includegraphics[width=\columnwidth]{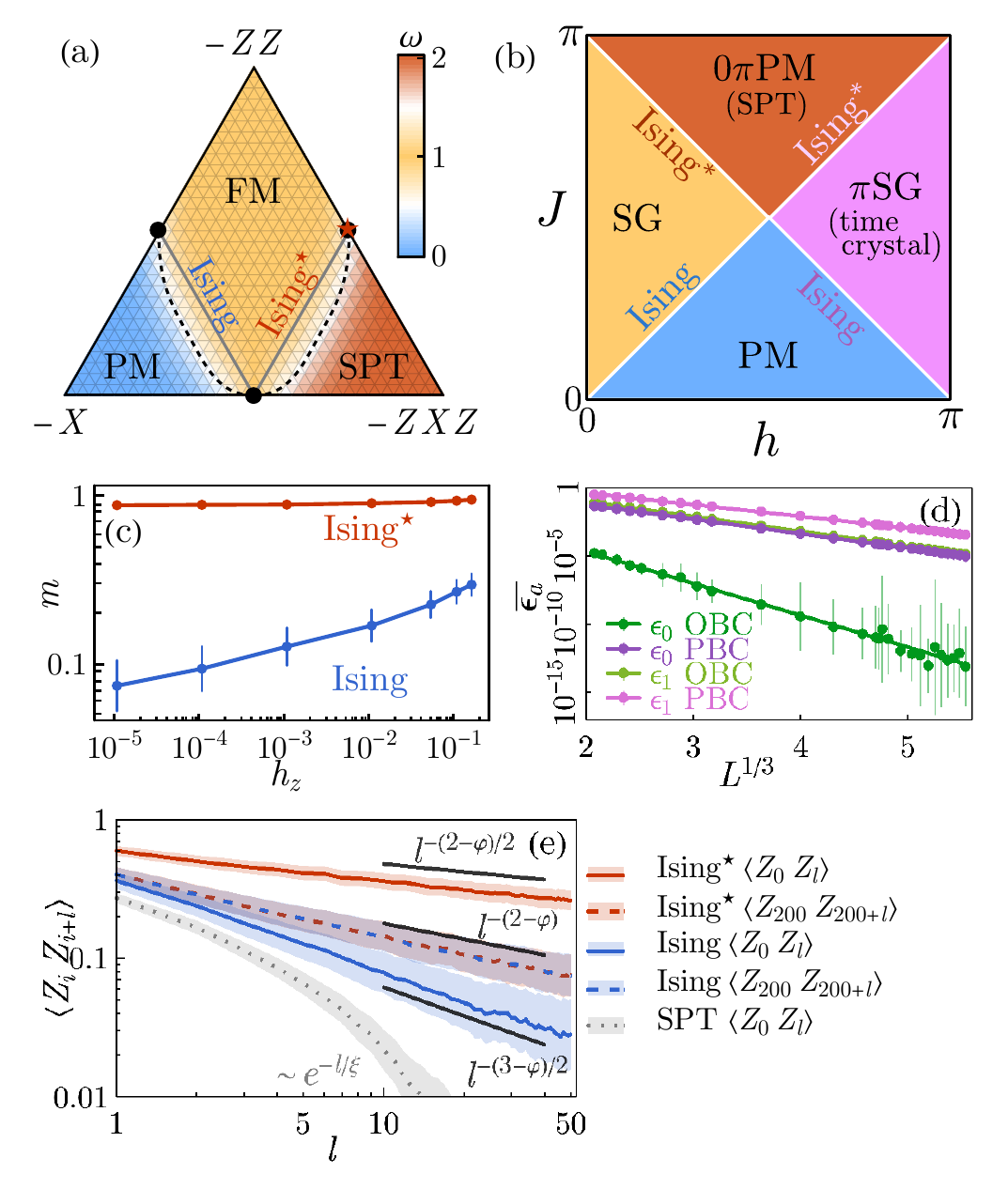}
\caption{ {\bf Random Ising$^{\star}$ transition} (a) Phase diagram of the random Ising Hamiltonian~\eqref{eq: Jhg model} for clean (solid lines) and disordered (dotted lines), showing the topological winding number $\omega$ for the dual fermionic description (see main text).
(b) Floquet phase diagram of eq.~\eqref{eqFloquet} which shows \emph{two} topologically non-trivial Ising$^\star$ transitions.
(c) Boundary magnetization under small Zeeman field, showing spontaneous magnetization at the Ising$^\star$ transition (red star in (a)). (d) Finite-size energy splitting of boundary spins at the Ising$^{\star}$ transition. (e) Spin-spin correlations involving bulk and boundary spins (averaged over $1.5\times10^5$ realizations), compared to theory predictions (solid black lines), where $\varphi$ is the golden ratio. Calculations are performed using the SBRG method on a 512-site spin chain.}
\label{fig: SBRG}
\end{center}
\end{figure}

{\bf Ising$^{\star}$ transition. } We consider the spin-$1/2$ chain
\begin{equation}\label{eq: Jhg model}
H = -\sum_{i}J_i Z_i Z_{i+1}-\sum_{i}h_i X_i -\sum_{i}g_i Z_{i-1}X_i Z_{i+1},
\end{equation}
where $X,Y,Z$ denote the Pauli matrices. The model has a $\dsZ_2$ spin-flip symmetry (generated by $P = \prod_i X_i$) and a time-reversal symmetry $\dsZ_2^T$ (acting as the complex conjugation $T=K$). Let us first consider the clean case, where the coefficients $J_i\equiv J$, $h_i \equiv h$ and $g_i \equiv g$ are site-independent. In this case, the $J,h,g \geq 0$ terms respectively drive the system towards ferromagnetic (FM), trivial paramagnetic (PM) and $\dsZ_2\times\dsZ_2^T$ symmetry protected topological (SPT) \cite{Suzuki71,Raussendorf01,Keating04,Son11,Verresen17} phases, the latter sometimes being called the cluster or Haldane SPT phase. The phase diagram is shown in \figref{fig: SBRG}(a), with the gray solid lines indicating Ising criticalities.

Although the FM-PM and FM-SPT transition are both described by the Ising conformal field theory (CFT), the time-reversal symmetry acts differently on the disorder operator, leading to different symmetry enriched CFTs (or gapless SPTs)~\cite{Scaffidi2017Gapless,Verresen2018Topology,Parker2018Topological,Verresen2019Gapless,Verresen2020Topology}.
To briefly review this, note that an Ising CFT has a unique local and a unique nonlocal scaling operator with scaling dimension $\Delta = 1/8$, commonly denoted by $\sigma$ and $\mu$, respectively. These are the order parameters of the nearby phases, i.e., $\sigma(n) \sim Z_n$ is the Ising order parameter, whereas the disorder operator $\mu(n)$ is the Kramers-Wannier-dual string order parameter of the symmetry-preserving phase. In the trivial PM, $\mu(n) \sim \prod_{j=-\infty}^n X_j$, whereas in the SPT phase, $\mu(n) \sim \prod_{j=-\infty}^n Z_{j-1}X_j Z_{j+1}= \cdots X_{n-2} X_{n-1} Y_n Z_{n+1}$ \cite{Smacchia11,Bahri14,Jones19,Verresen2019Gapless}. We see that the two Ising critical lines are distinguished by the discrete invariant $T \mu T = \pm \mu$ \cite{Verresen2019Gapless}. This means they must be separated by a phase transition. Indeed, in \figref{fig: SBRG}(a) they meet at a multicritical point where the central charge is $c=1$.

We refer to the non-trivial case, where the nonlocal bulk operator is charged $T \mu T = - \mu$, as Ising$^{\star}$. This supports a localized zero-energy edge state \cite{Verresen2019Gapless}. Intuitively, the edge of the Ising$^{\star}$ criticality spontaneously breaks the Ising $\mathbb Z_2$ symmetry. This unusual degenerate boundary fixed point is stable since $\mu$ is charged and hence cannot be used to disorder the boundary. The finite-size splitting of this edge mode is parametrically faster than the finite-size bulk gap $\sim 1/L$. In particular, if the model is dual to free-fermions (such as Eq.~\eqref{eq: Jhg model}) then the edge mode is exponentially-localized \cite{Verresen2018Topology} whereas with interactions, the splitting becomes $\sim 1/L^{14}$ \cite{Verresen2019Gapless}.

{\bf Random Ising$^{\star}$ transition. } We now study the fate of Ising$^{\star}$ upon disordering the system. The coefficients $J_i$, $h_i$ and $g_i$ in Eq.~\eqref{eq: Jhg model} are now independently drawn from power-law distributions $P(J)=(J/J_0)^{1/\Gamma}/(\Gamma J)$ for $J\in[0,J_0]$ (similarly for $P(h)$ and $P(g)$), where $\Gamma$ controls the width of the distribution in logarithmic scale. The limit $\Gamma \to 0$ would recover the clean case. We will take $\Gamma=1$, i.e., the uniform distribution.

In the presence of randomness, the Ising CFT flows towards the infinite-randomness fixed point ($\Gamma\to\infty$) \cite{Fisher1992Random,Fisher1995Critical}. We will explore the symmetry enriched infinite-randomness fixed point as the many-body localized counterpart of gapless SPT states. The disordered phase diagram is shown in \figref{fig: SBRG}(a), which is qualitatively unchanged from the clean case. This was obtained by mapping Eq.~\eqref{eq: Jhg model} to free fermions (using a Jordan-Wigner transformation) and using the transfer matrix method to determine the topological winding number $\omega$ \cite{Motrunich01}; in this case the PM, FM and SPT phases map to the trivial ($\omega=0$), Kitaev chain ($\omega=1$) and two Kitaev chains ($\omega=2$). In the original spin chain language, one can interpret $\omega$ as encoding the ground state degeneracy $2^\omega$ with open boundary conditions, which is $0$, $2$ and $4$, respectively.

Similar to the Ising CFT, the infinite-randomness Ising fixed point also has a local $\sigma$ and nonlocal $\mu$ scaling operator. While their scaling dimensions have changed ($\Delta^{\rm bulk} = 1-\varphi/2 \approx 0.191$ where $\varphi=\frac{1}{2}(1+\sqrt{5})$ is the golden ratio~\cite{Fisher1995Critical}), their lattice expressions are as before---indeed, the nearby gapped phases are still characterized by the same order parameters. We thus still have the bulk topological invariant $T \mu T = \pm \mu$, distinguishing two distinct symmetry-enriched infinite-randomness Ising fixed points, which we refer to as the Ising and Ising$^{\star}$. For the same reasons as before, we expect that the disordered Ising$^{\star}$ criticality has spontaneously-fixed boundary conditions. This would come with at least three physical fingerprints: (i) a nonzero spontaneous magnetization at the boundary, (ii) a degenerate edge mode whose finite-size splitting is parametrically smaller than the bulk gap, and (iii) spin-spin correlations near the boundary should have a boundary scaling dimension~\cite{McCoy1969Theory,Igloi1998Random} $\Delta_{\sigma}^\text{bdy}=1/2$ (or $0$) for free (or spontaneously-fixed) boundary condition, characterizing the Ising (Ising$^{\star}$) case.

We now test these predictions numerically. Because we will be interested in including interactions, we use
the spectrum bifurcation renormalization group (SBRG) method~\cite{You2016Entanglement,Slagle2016Disordered,Slagle2017Out-of-time-order,supmat}, which is a numerical real-space renormalization group approach that progressively transforms the original Hamiltonian $H$ to its diagonal form $H_\text{MBL}=\sum_{a}\epsilon_a\tau_a+\sum_{ab}\epsilon_{ab}\tau_a\tau_b+\cdots$ as a many-body localization (MBL) effective Hamiltonian~\cite{Abanin:2013lc,Huse:2014ec, Swingle:2013oy}, and constructs the (approximate) local integrals of motion $\tau_a$ of the MBL system in the form of Pauli strings. The approximation is asymptotically exact in the strong-disorder limit. The rescaled parameters $(\tilde{J},\tilde{h},\tilde{g})\equiv (J_0,h_0,g_0)^{1/\Gamma}$ are invariant under the renormalization group (RG) flow, and should be considered as tuning parameters. SBRG can be thought of as an implementation of the strong disorder real space renormalization group (RSRG)~\cite{Ma1979Random, Dasgupta1980Low-temperature,Fisher1992Random, Fisher1994Random, Fisher1995Critical, Motrunich2000Infinite-randomness} and its generalization to excited states (RSRG-X)~\cite{PekkerRSRGX,VoskAltmanPRL13,PhysRevLett.112.217204,QCGPRL} in operator space. While SBRG can be used to study MBL physics and excited states, in the following we focus on $T=0$ groundstate properties.

{\bf SBRG results. }
We focus on the  Ising$^{\star}$ transition at $(\tilde{J},\tilde{h},\tilde{g})=(1,0,1)$ (red star in \figref{fig: SBRG}(a)). 
We have verified~\cite{supmat} that in the bulk, the Ising$^{\star}$ transition flows to an infinite-randomness fixed point with dynamical scaling $l\sim(\log t)^2\sim(-\log \epsilon)^2$ that relates the length scale $l$ and the energy scale $\epsilon$~\cite{Fisher1995Critical}, and logarithmic scaling of the entanglement entropy~\cite{PhysRevLett.93.260602,Refael2009Criticality}. This is not surprising since with periodic boundary conditions, Ising and Ising$^{\star}$ are unitarily equivalent.

We now probe the boundary properties. To include the effect of interactions, we follow Ref.~\cite{Verresen2019Gapless} and add a generic $\mathbb Z_2 \times \mathbb Z_2^T$-symmetric boundary perturbation $H_V=-V(X_0Z_1Z_2+Z_{L-2}Z_{L-1}X_L)$, with $V$ a random variable ten times smaller than the bulk couplings. Microscopically, this perturbation can flip the boundary Ising spin. Nevertheless, if we study the boundary magnetization $m=\langle Z_0\rangle$ in response to a small Zeeman field $h_z$ applied along the $z$-axis, we find that it tends to a nonzero limit as $h_z \to 0$ (with $h_z$ smaller than the finite-size bulk gap, but larger than the groundstates splitting, see below), shown in \figref{fig: SBRG}(d). This is in contrast to the trivial Ising fixed point, where the boundary magnetization is known to vanish as $m(h_z)\sim1/|\log h_z|$~\cite{McCoy1969Theory}.

Thus the boundary is spontaneously magnetized in the Ising$^{\star}$ case despite the Hamiltonian~\eqref{eq: Jhg model} being symmetric. Schematically, on a finite system we have two spontaneously-fixed ferromagnetic (FM) ground states $\ket{\uparrow_L \uparrow_R}$ and $\ket{\downarrow_L \downarrow_R}$, where $L$ and $R$ denote the configurations of the left and right edge modes (note that these are split from $\ket{\uparrow_L \downarrow_R}$ and $\ket{ \downarrow_L \uparrow_R}$ by the critical bulk penalizing antiferromagnetic states)~\cite{Scaffidi2017Gapless,Verresen2019Gapless}. The above perturbation $H_V$ can couple these FM states at second order in $V$, which should lead to a finite-size splitting. The claim that we have a ground state degeneracy is only meaningful if this splitting is smaller than the bulk finite-size gap. To confirm this, we arrange the energy coefficients $\epsilon_a$ obtained from SBRG in the ascending order $\epsilon_0<\epsilon_1<\cdots$, and focus on the lowest two. For the  Ising$^{\star}$ transition with open boundary condition (OBC), $\epsilon_0$ characterizes the smallest energy splitting between $\ket{\uparrow_L\uparrow_R}\pm\ket{\downarrow_L\downarrow_R}$ whereas $\epsilon_1$ characterizes the bulk excitation gap. As shown in \figref{fig: SBRG}(e), both splittings $\overline{\epsilon_0}$ and $\overline{\epsilon_1}$ follow $\overline{\epsilon_a}\sim \exp({-\alpha_a L^{1/3}})$ but with different exponents $\alpha_0= 5.4\pm0.6$ and $\alpha_1=2.51\pm0.02$, i.e., $\overline{\epsilon_0} \approx \overline{\epsilon_1}^2$. The finite-size splitting $\overline{\epsilon_0}$ of the symmetry-protected edge modes decays significantly faster with the system size $L$ compared to $\overline{\epsilon_1}$. This provides a quantitative distinction between the topological edge modes and the bulk excitations. To further verify this interpretation, we switch to the periodic boundary condition (PBC), the fast-decaying topological splitting disappears and the smallest splitting decays with the ``bulk'' exponent as $\alpha_0=2.45\pm0.02$.

The Ising and  Ising$^{\star}$ states can be further distinguished by their average boundary-bulk spin-spin correlation functions $\overline{\langle Z_0 Z_l\rangle}$, which decay as $\sim1/l^{\Delta_{\sigma}^\text{bdy}+\Delta^\text{bulk}}$, where $\Delta_\sigma^\text{bdy}$ ($\Delta^\text{bulk}$) is the boundary (bulk) scaling dimension of the Ising order parameter mentioned before.
We thus predict
\begin{equation}\label{eq: bdy-blk exponent}
\overline{\langle Z_0 Z_l\rangle} \sim\left\{\begin{array}{cc}l^{-(3-\varphi)/2}\approx l^{-0.69} & \text{Ising},\\l^{-(2-\varphi)/2}\approx l^{-0.19} & \text{ Ising$^{\star}$}.\end{array}\right.
\end{equation}
In \figref{fig: SBRG}(f), we find that the boundary-bulk correlation follows $\overline{\langle Z_0 Z_l\rangle}\sim l^{-(0.67\pm0.08)}$ for Ising and $ l^{-(0.20\pm0.02)}$ for  Ising$^{\star}$, which matches \eqnref{eq: bdy-blk exponent} within error bars.
We also checked that the bulk-bulk correlation $\overline{\langle Z_i Z_{i+l}\rangle}\sim l^{-(0.42\pm0.05)}$ decays with the expected exponent $2\Delta^\text{bulk}=2-\varphi\approx 0.38$ for both Ising and Ising$^{\star}$ transitions.

{\bf Symmetry-enriched random singlet phase. } The Ising$^{\star}$ transition provides a clear example of symmetry-enriched random quantum critical point, with stretched-exponentially localized edge modes. It is natural to ask whether this notion can be extended to random {\em critical phases}, and whether the topological edge modes can be made {\em exponentially} localized despite the absence of a bulk gap. Here, we answer both questions in the positive, by introducing a symmetry-enriched random singlet phase.

In order to obtain a critical phase in one dimension, we consider a system with charge conservation and particle-hole symmetry. For concreteness, we will focus on the random antiferromagnetic spin-$1/2$ XXZ spin chain $H_{\rm A} = \sum_i J_i (X^A_{i} X^A_{i+1} + Y^A_{i} Y^A_{i+1} + \Delta_i Z^A_{i} Z^A_{i+1} )$, with $J_i>0$ and $0<\Delta_i<1$ random couplings specified later. It has a symmetry group $G_A=U(1) \rtimes \dsZ_2^A$ with the $\dsZ_2^A$ spin flip generated by $\prod_i X^A_i$, while the $U(1)$ part corresponds to $\sum_i Z^A_i$ conservation. For uniform couplings, this spin chain is in a Luttinger liquid phase; while for random couplings, its low energy properties can be captured by a real-space renormalization group (RSRG) procedure very similar to the SBRG approach above (but restricted to the groundstate). The random XXZ spin chain forms a {\em random singlet phase}~\cite{Fisher1994Random}, where the groundstate is asymptotically made of non-crossing pairs of singlets of all ranges, with quantum critical properties similar to the random Ising transition (which itself can be thought of as a random singlet state of Majorana fermions). In particular, the entanglement entropy grows logarithmically with effective central charge $c_{\rm eff}=\log 2$ \cite{Refael07,Refael2009Criticality}, and the gap closes stretched-exponentially with system size (dynamical exponent $z=\infty$). 

To obtain a topological random singlet phase, we use the {\em decorated domain walls} construction~\cite{DecoratedDW} to ``twist'' the random XXZ chain. To that effect, we introduce another spin species $B$, with Ising symmetry $G_B = \dsZ_2^B$, with Hamiltonian $H_B= - \sum_{i} X^{B}_i  + g_B Z^{B}_i Z^{B}_{i+1}.$ We take $g_B \ll 1$ so that the $B$ spins are disordered, deep into a quantum paramagnetic phase. 
We then couple the two models by attaching charges of the $G_B = \dsZ_2^B$ symmetry to the domain walls of the $A$ spins. This is achieved by the unitary transformation $U = \prod_{{\rm DW}(A)} (-1)^{(1-Z^B_i)/2}$, where the product runs over all the domain walls of the $A$ spins in the $Z$ basis, with $U^2=1$. After unitary rotation (``twist'') of $H_A+H_B+V$, we find
\begin{align}\label{eq:XXZmodel}
H &= \sum_{i} J_i\left[ Z^B_{i-1}(X^A_{i} X^A_{i+1} + Y^A_{i} Y^A_{i+1}) Z^B_{i+1} + \Delta_i Z^A_{i} Z^A_{i+1} \right] \notag \\
- &\sum_{i} Z^A_i X^{B}_i Z^A_{i+1} + g_B Z^{B}_i Z^{B}_{i+1} + V',
\end{align}
where $V'=UVU$ represents arbitrary small perturbations that preserves the $G_A \times G_B$ symmetry. Following the terminology of Ref.~\cite{Scaffidi2017Gapless,Parker2018Topological}, we refer to Eq.~\eqref{eq:XXZmodel}  and $H_A+H_B+V = UHU$ as the gSPT and gTrivial (gapless, topologically trivial) Hamiltonians, respectively.  

For periodic boundary conditions, $H$ is unitarily related to $H_A + H_B$ plus perturbations, and thus corresponds to random singlet $A$ spins coupled to the gapped paramagnetic $B$ spins. Nevertheless, the two models are topologically distinct. Like Ising and Ising$^\star$ above, they can be distinguished by the charges of nonlocal scaling operators. In fact, since now there are additional gapped degrees of freedom, one can consider a string order parameter with long-range order: in the trivial case $H_A+H_B$ this is $  \cdots X^B_{j-2} X^B_{j-1} X^B_j$ whereas in the topological case $H$ it is $ \cdots X^B_{j-2} X^B_{j-1} X^B_j Z^A_{j+1}$. In the latter case, this string order parameter for the gapped $B$ variables is charged under $G_A$. This discrete invariant shows that we have two distinct symmetry-enriched versions of the same underlying infinite-randomness fixed point. Relatedly, for open boundary conditions, we have $H = J_0 \Delta_0 Z_0^A Z_1^A + Z^A_0  X_0^B Z_1^B + Z_0^B Z_1^B + \dots$, and in the absence of additional perturbations ($V=0$), we see that $[Z_0^A ,H]=0$, providing an exact edge mode. 

\begin{figure}[t!]
\begin{center}
\includegraphics[width=\columnwidth]{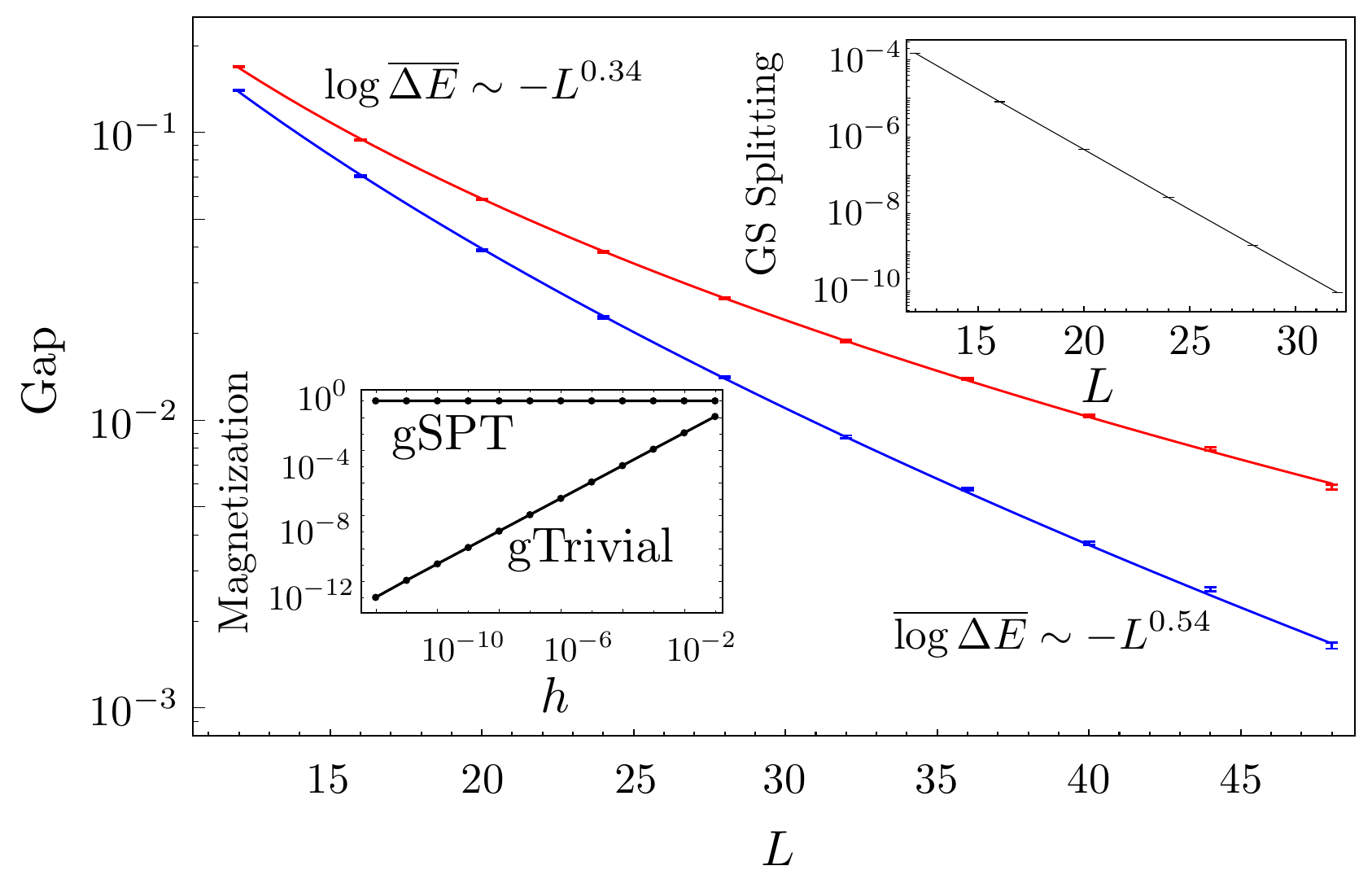}
\caption{ {\bf Symmetry-enriched random singlet phase.} DMRG results on eq.~\eqref{eq:XXZmodel} including various perturbations~\cite{supmat}. 
Fits of the typical and average finite size gaps, showing a scaling compatible with the random-singlet $z=\infty$ scalings $\Delta E_{\rm typical} \equiv {\rm e}^{\overline{\log \Delta E}} \sim {\rm e}^{-\sqrt{L}}$ and $\overline{\Delta E} \sim {\rm e}^{-L^{1/3}}$. {\it Top-right Inset:} the splitting between the two ground states vanishes exponentially with system size, indicating exponentially-localized edge modes.  {\it Bottom-left Inset:} spontaneous boundary magnetization in the presence of a small symmetry-breaking magnetic field $h$. }
\label{fig:DMRG}
\end{center}
\end{figure}

 Going away from this special limit, we expect exponentially-localized topological edge modes to be protected by the finite gap of the $B$ spins, as in the clean case~\cite{Scaffidi2017Gapless,Parker2018Topological}. We confirmed numerically the presence of exponentially localized edge modes coexisting with bulk random singlet criticality using density-matrix renormalization group (DMRG)~\cite{White93,ITensor} techniques (Fig.~\ref{fig:DMRG}), including generic symmetry-preserving perturbations~\cite{supmat}.

{\bf Floquet Ising$^{\star}$ criticality.} To close this letter, we illustrate how such novel universality classes emerge naturally in the context of periodically driven (Floquet) systems. We focus on the driven quantum Ising chain characterized by the single-period evolution (Floquet) operator~\cite{PhysRevLett.116.250401}
\begin{equation} \label{eqFloquet}
F = {\rm e}^{- \frac{i}{2} \sum_i J_i Z_i Z_{i+1} + \dots} {\rm e}^{- \frac{i}{2} \sum_i h_i X_i + \dots}
\end{equation}
where the dots represent small but arbitrary interactions preserving the ${\mathbb Z}_2$ symmetry $G=\prod_i X_i$. For strong enough disorder, this system admits four dynamical phases protected by
MBL~\cite{PhysRevLett.116.250401}. In addition to the familiar paramagnetic (PM) and spin glass (SG) Ising phases, there are two more phases called $\pi$-SG (a.k.a. time crystal~\cite{PhysRevLett.116.250401,PhysRevLett.117.090402,PhysRevB.94.085112,Zhang:2017aa, 2019arXiv191010745K}) and $0 \pi$ PM (a non-trivial SPT phase); see Fig.~\ref{fig: SBRG}(b). This phase structure is due to an emergent ${\mathbb Z}_2$ symmetry inherited from time translation symmetry. The transitions between those phases have been argued to be in the random Ising universality class~\cite{PhysRevLett.118.030401,Berdanier9491} (ignoring potential instabilities towards thermalization in the presence of interactions~\cite{2020arXiv200809113M,2020arXiv200808585S,2020arXiv201010550W}). Here we note that the transitions out of the $0 \pi$ PM are actually in the random Ising$^{\star}$ universality class described above, protected by ${\mathbb Z}_2 \times {\mathbb Z}_2$ symmetry (one of the ${\mathbb Z}_2$'s being emergent). This is because the $0 \pi$ PM is closely related to the ${\mathbb Z}_2 \times {\mathbb Z}_2$ equilibrium SPT~\cite{PhysRevLett.116.250401, vonKeyserlingk2016a,Else16,Potter16,Harper16}. We find exponentially localized edge modes at the transitions separating the $0 \pi$ PM to either the SG or $\pi$-SG, which are protected due to the disorder operator $\mu$ for the critical $\mathbb Z_2$ symmetry again being charged with respect to the second $\mathbb Z_2$ symmetry, as detailed in the supplemental material~\cite{supmat}. (The edge mode localization is exponential as in the random singlet phase above, as the protecting symmetry is $\mathbb Z_2 \times \mathbb Z_2$ instead of $\mathbb Z_2 \times \mathbb Z_2^T$.)

{\bf Discussion.} We have demonstrated the existence of symmetry-enriched infinite-randomness fixed points with robust topological edge modes coexisting with all the characteristics of strong disorder quantum criticality. In particular, we have shown that the paradigmatic random Ising critical point and XXZ random singlet phase come in topologically distinct versions in the presence of an additional ${\mathbb Z}_2^T$ or ${\mathbb Z}_2$ symmetry. The topological edge modes couple non-trivially to gapless bulk fluctuations, leading to anomalous boundary critical behavior. We expect our findings to extend to essentially all known strong- and infinite-randomness critical points: finding examples of symmetry-enriched random critical points in 2+1d~\cite{Motrunich2000Infinite-randomness,BMinprep} and 3+1d represents an interesting direction for future works. It would also be interesting to investigate the consequences of our results for dynamical properties~\cite{PhysRevLett.122.240605, Berdanier9491,2019arXiv191205546K,PhysRevLett.124.206803}.

\begin{acknowledgments}
{\it Acknowledgments.} We thank Sid Parameswaran, Brayden Ware and Shang Liu for useful discussions. We also thank Nick G. Jones, Ryan Thorngren, Daniel Parker, Frank Pollmann, and Thomas Scaffidi for collaborations on related matters. V. Khemani, R. Verresen and R. Vasseur are especially grateful to William Berdanier for early collaboration and discussion on Floquet criticality. The MPS-based DMRG simulations were performed using the ITensor Library~\cite{ITensor}, and SBRG simulations were performed using SBRG(2.0)~\cite{SBRG_algorithm}. This work was supported by the US Department of Energy, Office of Science, Basic Energy Sciences, under Early Career Award Nos. DE-SC0019168 (R. Vasseur) and DE-SC0021111 (V. Khemani), the Alfred P. Sloan Foundation through a Sloan Research Fellowship (R. Vasseur and V. Khemani), the Harvard Quantum Initiative Postdoctoral Fellowship in Science and Engineering (R. Verresen), a grant from the Simons Foundation (\#376207, Ashvin Vishwanath) (R. Verresen), and a startup fund from UCSD (HY~Hu and YZ~You). 

\end{acknowledgments}

\bibliography{ref}

\bigskip

\onecolumngrid
\newpage



\section*{Supplementary material}

# Supplemental material for "Topological and symmetry-enriched random quantum critical points"

Carlos M. Duque and Romain Vasseur
*Department of Physics, University of Massachusetts, Amherst, Massachusetts 01003, USA*

Hong-Ye Hu and Yi-Zhuang You
*Department of Physics, University of California, San Diego, CA 92093, USA*

Vedika Khemani
*Department of Physics, Stanford University, Stanford, CA 94305, USA*

Ruben Verresen
*Department of Physics, Harvard University, Cambridge, MA 02138, USA*

## I. SPECTRUM BIFURCATION RENORMALIZATION GROUP

In this appendix, we briefly summarize the spectrum bifurcation renormalization group (SBRG) method[1–3]. The basic idea of SBRG is to progressively identify conserved quantities by block diagonalization, and treat the off-diagonal term within second order perturbation theory. The general form of a qubit model Hamiltonian can be written as

$$H = \sum_{[\mu]} h_{[\mu]} \sigma^{[\mu]}, \qquad (1)$$

where $\sigma^{\mu_i}$ ($\mu_i = 0,1,2,3$) are Pauli matrices acting on the $i$th site, and $\sigma^{[\mu]} \equiv \sigma^{\mu_1} \otimes \sigma^{\mu_2} \otimes \sigma^{\mu_3} \cdots$ denotes a string of Pauli operators. The real coefficients $h_{[\mu]}$ are drawn from some random distribution. To progressively diagonalize the Hamiltonian in Eq.(1), we first find the leading energy scale term

$$H_0 = h_{[\mu]_{\max}} \sigma^{[\sigma]_{\max}}, \qquad (2)$$

where $|h_{[\mu]_{\max}}|$ has the largest value among all the coefficients. We then block diagonalize term $\sigma^{[\mu]_{\max}}$ by a Clifford rotation $R$,

$$R^\dagger \sigma^{[\mu]_{\max}} R = \sigma^{3[0\cdots 0]}, \qquad (3)$$

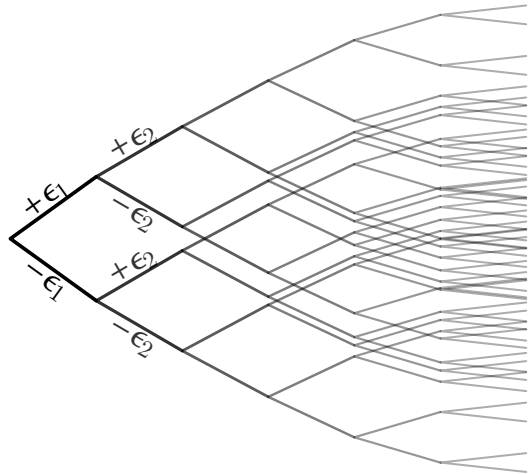

FIG. 1: **Spectrum bifurcation renormalization group.** Spectrum of energy generated by SBRG: at each RG steps, the RG bifurcates into a high- and low-energy branch, corresponding to $\tau_i = \pm 1$ respectively. Each many-body eigenstate corresponds to a leaf of the tree, uniquely labelled by the values $\{\tau_i = \pm 1\}$.

and all the other terms will be rotated by the same operator $R$, such that $H \to R^\dagger H R$. As we block diagonalize the leading energy term, the many-body spectrum bifurcates to high energy $E \sim |h_{[\mu]_{\max}}|$ sector, or to low energy $E \sim -|h_{[\mu]_{\max}}|$ sector. We now reduce the other leftover Hamiltonian to either sector. The Hamiltonian can be separated into the following three terms:

$$H = H_0 + \Delta + \Sigma, \tag{4}$$

where $\Delta$ are the terms that commute with the leading term $H_0$, and $\Sigma$ are the terms that anti-commute with $H_0$. We will renormalize the off-diagonal term $\Sigma$ using second order perturbation theory, $H \to S^\dagger H S$, with $S = \exp\left(-H_0 \Sigma / 2 h_{[\mu]_{\max}}\right)$. The effective Hamiltonian reads

$$H = H_0 + \Delta + \frac{1}{2 h_{[\mu]_{\max}}^2} H_0 \Sigma^2 + \mathcal{O}\left(\frac{1}{2 h_{[\mu]_{\max}}^3}\right). \tag{5}$$

And we will only keep terms generated up to 2nd order in perturbation theory. We then repeat this procedure at $\mathcal{O}(N)$ times to fully diagonalize the original many-body Hamiltonian. At the end of RG flow, we will obtain the effective Hamiltonian

$$H_{\text{eff}} = \sum_i \epsilon_i \tau_i + \sum_{i<j} \epsilon_{ij} \tau_i \tau_j + \cdots, \tag{6}$$

where $\tau_i = \pm 1$ are the emergent conserved quantities.

The whole RG transformation can be summarized as a unitary transformation,

$$U_{\text{RG}} = R_1 S_1 R_2 S_2 \cdots = \prod_i R_i S_i, \tag{7}$$

and $H \to U_{\text{RG}}^\dagger H U_{\text{RG}} \approx H_{\text{eff}}$. The eigenstates can be labeled by a vector of conserved quantities, i.e. $|\tau\rangle = |\tau_1 = 1, \tau_2 = -1, \cdots\rangle$. We can effectively calculate the expectation value of any physical observable $\hat{O}$ in the $\{\tau_i\}$ basis as

$$\langle \psi | \hat{O} | \psi \rangle = \langle \psi | U_{\text{RG}} U_{\text{RG}}^\dagger \hat{O} U_{\text{RG}} U_{\text{RG}}^\dagger | \psi \rangle = \langle \tau | \hat{O}_{\text{eff}} | \tau \rangle. \tag{8}$$

The entanglement entropy can also be efficiently calculate using the stabilizer formalism[1,4].

We investigated the bulk property of Ising* transition as a sanity check. In Fig.2 (a) show the entanglement entropy $S_E = -\text{tr}\rho_A \log \rho_A$ grows logarithmically with the size $L_A$ of the entanglement region following $S_E(L_A) = (c_{\text{eff}}/3) \log L_A$[5]. We observe the "effective central charge" $c_{\text{eff}} = (0.51 \pm 0.02) \log 2$, which is consistent with the expectation $c_{\text{eff}} = (1/2) \log 2$ for the infinite-randomness fixed point in the Ising universality class. Another signature of the infinite-randomness fixed point is the dynamical scaling $l \sim (\log t)^2 \sim (-\log \epsilon)^2$ that relates the length scale l and the energy scale $\epsilon$[6]. We check this by examining the relation between the length $l_a$ of the Pauli string $\tau_a$ and its corresponding energy coefficient $\epsilon_a$ in $H_{\text{MBL}}$. Fig. 2 (b) confirms the scaling $\overline{\log \epsilon_a} \sim -\sqrt{l_a}$. These evidences justify that the disordered Ising* does flow to an infinite-randomness fixed point.

## II. DMRG ANALYSIS OF THE SYMMETRY-ENRICHED RANDOM SINGLET PHASE

In order to study the stability of the symmetry-enriched random singlet phase discussed in the main text, we included perturbations to take the $B$ spins away from integrability, to further couple the $A$ and $B$ spins, and we also added generic symmetry-preserving terms near the edges. In the following, we take $J_i$ and $\Delta_i$ to be drawn from uniform random distributions, between $[0.1, 1.0]$ and $[0.3, 0.7]$, respectively; $g_B = 0.3$, and $V = -\sum_i 0.1 X_i^B X_{i+1}^B + 0.1 Z_i^A X_i^B Z_{i+1}^A$ where the first term breaks integrability of the $B$ spins while the second term couples $A$ and $B$ spins before the unitary twist. We also add small fields $0.2(X_0^A + X_N^A) + (X_0^B + X_N^B)$ that preserve the $\mathbb{Z}_2^A \times \mathbb{Z}_2^B$ symmetry. (Note that those boundary terms break the $U(1)$ $\sum_i Z_i^A$ conservation at the boundary, but that symmetry is only required to tune the bulk to criticality, and does not play a role in protecting the edge modes.)

Because of the non-Abelian symmetry $G_A$ of the model, the SBRG approach used to study eq. 1 in the main text does not apply here[15]. In order to establish the presence of topological edge modes and study their interplay with the coexisting quantum critical bulk fluctuations, we compute the low-energy spectrum using density-matrix renormalization group (DMRG) techniques[7,8]. While DMRG cannot access system sizes as large as SBRG, it has the advantaged of being essentially numerically exact for reasonably small system sizes ($L \lesssim 50$ before numerical

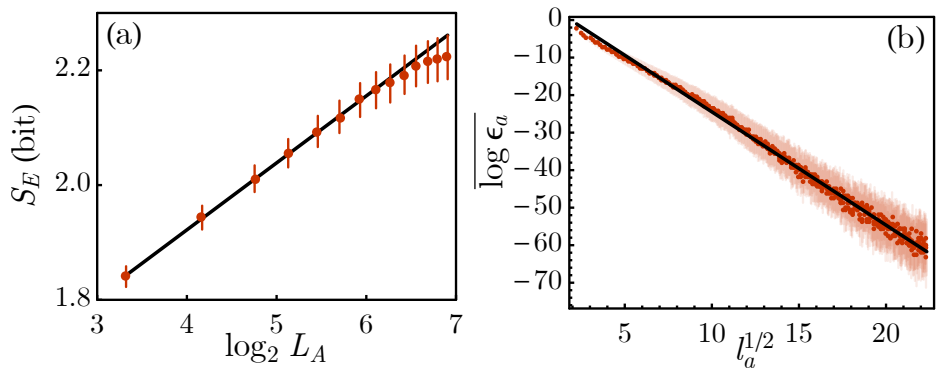

FIG. 2: **Bulk criticality of the Ising$^\star$ fixed point** (a) Entanglement entropy scaling and (b) dynamical scaling (collected over $10^4$ realizations) at the Ising$^\star$ fixed point.

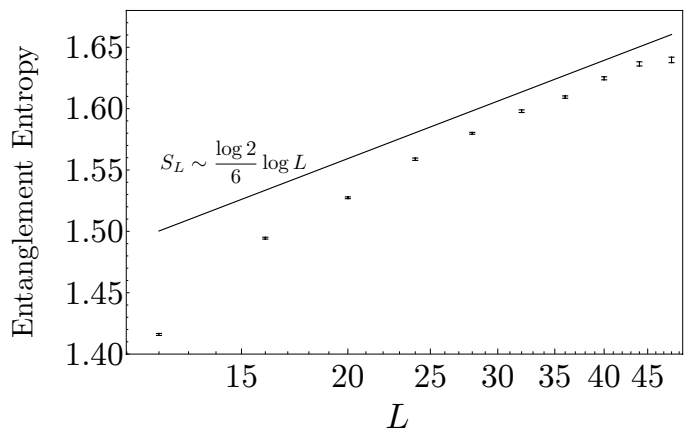

FIG. 3: **Entanglement scaling in the symmetry-enriched random singlet phase.** Bipartite entanglement entropy *vs* system size in the groundstate of the symmetry-enriched random singlet phase, computed using DMRG. The numerical results are consistent with the random singlet phase prediction $S \sim \log 2/6 \log L$ (solid lines).

instabilities make the variational sweeps unpractical). We used the ITensor library[8] to perform DMRG simulations of the disordered symmetry-enriched random singlet Hamiltonian, using bond dimension $\chi \sim 500$. We considered system sizes $L = 2N = 12, 16, 20, \ldots, 48$, where $N$ is the number of spins per species $A, B$. The number of disordered realizations ranged from $\sim 6000$ (for the largest system size $L = 48$) to $3 \times 10^4$. We found that a large number of disorder realizations was necessary to observe the expected random singlet quantum critical behavior. However, the topological edge modes can be observed on individual disorder realizations.

We find an exponentially small in system size splitting between the two groundstates, $|\uparrow_L\downarrow_R\rangle \pm |\downarrow_L\uparrow_R\rangle$, where $R$ and $L$ denote the configurations of the right and left edge modes, as in the main text. The orientation of the edge modes can be detected from the spontaneous boundary magnetization in each eigenstate upon adding a small magnetic field. We observe numerically that those groundstates are separated from the first excited states $|\uparrow_L\uparrow_R\rangle \pm |\downarrow_L\downarrow_R\rangle$: critical bulk fluctuations separate ferro- and antiferromagnetic alignments of the edge modes, leading to a 2-fold groundstate degeneracy. Coexisting with these topological edge modes, we find bulk quantum critical properties characteristic of a random singlet phase. The typical bulk gap closes as $\Delta E_{\rm typical} \equiv e^{\overline{\log \Delta E}} \sim e^{-\sqrt{L}}$, with as in the Ising case, a broad gap distribution leading to a different scaling for the average $\overline{\Delta E} \sim e^{-L^{1/3}}$. This stretched-exponential gap closing is hallmark of an infinite randomness quantum critical point, and is apparent in our DMRG results as shown in the main text. We find other signatures of random-singlet criticality: in particular, the entanglement entropy (using open boundary conditions, with the entanglement cut in the middle of the system) grows logarithmically with system size $L$ (Fig. 3), with a prefactor compatible with the prediction $S \sim \frac{\log 2}{6} \log L$[9].

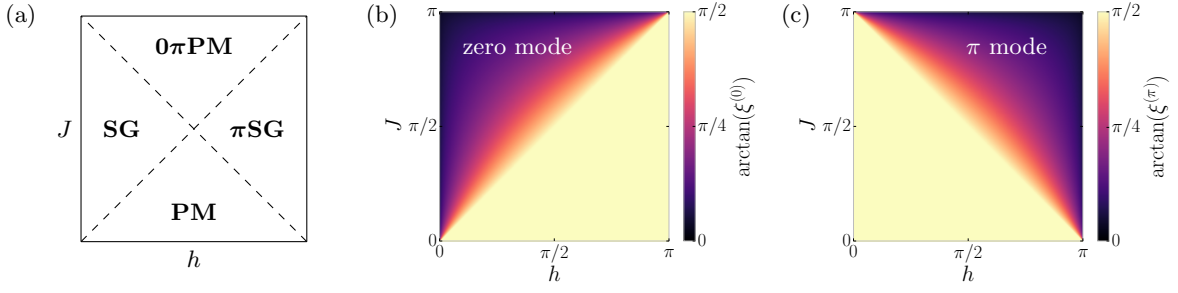

FIG. 4: **Topological edge modes at Floquet criticality.** (a) Schematic representation of the Floquet phase diagram of eq. (11). (b) The localization length of the zero mode; note that it remains <u>finite</u> (purple region) at the critical line between SG and $0\pi$PM (c) Similarly, the localization length for the $\pi$ mode remains finite at the transition between $\pi$SG and $0\pi$PM.

### III.  FLOQUET ISING$^\star$ CRITICALITY

In this appendix, we analyze the edge mode structure of the Floquet transverse Ising chain. We focus on the non-interacting limit for simplicity, and also consider uniform couplings as most features we want to illustrate are already apparent in this limit. Disorder is needed to many-body localize the phases in the presence of interactions, and to make the notion of Floquet phase structure meaningful away from the non-interacting limit. (Without disorder, Floquet systems typically heat up indefinitely and thermalize to infinite temperature.) Let us work in the Majorana representation with a pair of Majorana fermions $\gamma_j, \tilde{\gamma}_j$ per site $j$, and define

$$H_\alpha \equiv \frac{i}{2} \sum_n \tilde{\gamma}_n \gamma_{n+\alpha}. \tag{9}$$

Under time evolution, we have $\gamma_n(t) = e^{iH_\alpha t}\gamma_n e^{-iH_\alpha t} = \cos(t)\gamma_n - \sin(t)\tilde{\gamma}_{n-\alpha}$, and $\tilde{\gamma}_n(t) \equiv e^{iH_\alpha t}\tilde{\gamma}_n e^{-iH_\alpha t} = \cos(t)\tilde{\gamma}_n + \sin(t)\gamma_{n+\alpha}$. If we introduce the unitaries

$$U_0 \equiv e^{-ihH_0} \qquad \text{and} \qquad U_1 \equiv e^{iJH_1}, \tag{10}$$

we have, after a Jordan-Wigner transformation

$$F = U_1 U_0 = e^{-\frac{i}{2}\sum_i JZ_i Z_{i+1}} e^{-\frac{i}{2}\sum_i hX_i}, \tag{11}$$

which is indeed the Floquet operator of the driven Ising chain.

Under Floquet stroboscopic evolution, we have

$$F^\dagger \gamma_n F = \begin{cases} \cos(J)\left(\cos(h)\gamma_n - \sin(h)\tilde{\gamma}_n\right) + \sin(J)\left(\cos(h)\tilde{\gamma}_{n-1} + \sin(h)\gamma_{n-1}\right) & \text{if } n > 1 \\ \cos(h)\gamma_1 - \sin(h)\tilde{\gamma}_1 & \text{if } n = 1 \end{cases} \tag{12}$$

$$F^\dagger \tilde{\gamma}_n F = \begin{cases} \cos(J)\left(\cos(h)\tilde{\gamma}_n + \sin(h)\gamma_n\right) - \sin(J)\left(\cos(h)\gamma_{n+1} - \sin(h)\tilde{\gamma}_{n+1}\right) & \text{if } n < N \\ \cos(h)\tilde{\gamma}_N + \sin(h)\gamma_N & \text{if } n = N. \end{cases} \tag{13}$$

From these expressions, we can construct the exact 0 and $\pi$ modes for a half-infinite chain[10]:

$$\gamma_L^{(0)} \equiv \sum_{n=1}^\infty \left(\frac{\tan(h/2)}{\tan(J/2)}\right)^{n-1} \left(\cos(h/2)\gamma_n - \sin(h/2)\tilde{\gamma}_n\right), \tag{14}$$

$$\gamma_L^{(\pi)} \equiv \sum_{n=1}^\infty \left(-\frac{1}{\tan(h/2)\tan(J/2)}\right)^{n-1} \left(\sin(h/2)\gamma_n + \cos(h/2)\tilde{\gamma}_n\right). \tag{15}$$

Here the labels 0 and $\pi$ refer to the quasi-energies of those modes, that is, $F^\dagger \gamma_L^{(0)} F = \gamma_L^{(0)}$, and $F^\dagger \gamma_L^{(\pi)} F = e^{i\pi}\gamma_L^{(\pi)}$.

Those edge modes are only meaningful when they are localized and normalizable. We plot the localization length of those modes as a function of $J$ and $h$, and compare it to the known phase diagram of this model[11] (Fig. 4). As expected, all four phases can be distinguished by their edge mode content: the paramagnet (PM) has no edge mode, the spin glass (SG) Ising symmetry-broken phase has a 0-Majorana mode, the $\pi$-spin glass ($\pi$SG) has a $\pi$-Majorana

mode, and finally, the 0πPM-paramagnet (0πPM) has both a 0 and a π-Majorana mode. Remarkably, we find that the localization length of the zero mode remains finite at the critical line between SG and 0πPM. Similarly, the localization length for the π mode remains finite at the transition between πSG and 0πPM. In other words, the transitions out of the 0πPM phase are in the Ising$^\star$ universality class. While we focused on the clean (uniform) case for simplicity, we have checked numerically that this carries over to the disordered case, and that both transition lines out of the 0πPM support exponentially localized edge modes. This Ising$^\star$ critically is protected by a $\mathbb{Z}_2 \times \mathbb{Z}_2$ symmetry, where one of the $\mathbb{Z}_2$'s is emergent and inherited from time translation symmetry. Indeed, the SG-0πPM transition will have an Ising disorder operator $\mu$ (for the critical $\mathbb{Z}_2$ Ising symmetry) which is charged under the emergent $\mathbb{Z}_2$ time-translation symmetry (this protects the edge mode for the same reasons as the $\mathbb{Z}_2 \times \mathbb{Z}_2^T$-symmetric case studied in the main text); analogously, the Ising disorder operator $\mu$ for the πSG-0πPM transition corresponds to the time translation $\mathbb{Z}_2$ which is now charged under the explicit Ising symmetry $\prod_n X_n$. Disorder is essential to meaningfully discuss the phase structure of Floquet systems in the presence of interactions, so the relevant criticality in general is the random Ising$^\star$ universality class discussed in the main text.



\end{document}